# Finding Bugs with Specification-Based Testing is Easy!

Janice Chin[a] and David J. Pearce[a]

a    Victoria University of Wellington, New Zealand

**Abstract**    Automated specification-based testing has a long history with several notable tools having emerged. For example, QuickCheck for Haskell focuses on testing against user-provided properties. Others, such as JMLUnit, use specifications in the form of pre- and post-conditions to drive testing. An interesting (and under-explored) question is how *effective* this approach is at finding bugs in practice. In general, one would assume automated testing is less effective at bug finding than static verification. *But, how much less effective?* To shed light on this question, we consider automated testing of programs written in Whiley — a language with first-class support for specifications. Whilst originally designed with static verification in mind, we have anecdotally found automated testing for Whiley surprisingly useful and cost-effective. For example, when an error is detected with automated testing, a counterexample is always provided. This has motivated the more rigorous empirical examination presented in this paper. To that end, we provide a technical discussion of the implementation behind an automated testing tool for Whiley. Here, a key usability concern is the ability to parameterise the input space, and we present novel approaches for references and lambdas. We then report on several large experiments investigating the tool's effectiveness at bug finding using a range of benchmarks, including a suite of 1800+ mutants. The results indicate the automated testing is effective in many cases, and that sampling offers useful performance benefits with only modest reductions in bug-finding capability. Finally, we report on some real-world uses of the tool where it has proved effective at finding bugs (such as in the standard library).

**ACM CCS 2012**

- **Software and its engineering** → **Software testing and debugging**;
- **Theory of computation** → *Automated reasoning*;

# The Art, Science, and Engineering of Programming

**Finding Bugs with Specification-Based Testing is Easy!**

# 1 Introduction

Automated testing is an increasingly important approach to program correctness [36, 49, 102]. One challenge is to generate sufficiently many inputs to thoroughly test the program in question. For example, *American Fuzzy Lop* [103] starts from a set of input seeds which it gradually mutates — an approach widely adopted [63, 95]. Other related techniques include *symbolic execution* [87], *feedback-directed testing* [72, 73] and *mutation testing* [42]. QuickCheck is another popular tool where tests are randomly generated based on user-provided properties [28]. QuickCheck has been demonstrated on industrial-scale projects (e.g., 1MLOC of C for Volvo [47]) and, whilst originally developed for Haskell, has subsequently been implemented for other languages including Erlang, Java and C [6, 29].

Functional specifications offer significant benefits for random testing [5, 88] and much work exists on *specification-based testing* [1, 9, 16, 18, 20, 24, 35, 36, 54, 62, 64, 65, 70, 71, 81, 91, 104]. Whittaker noted "Without a specification, testers are likely to find only the most obvious bugs" [99]. Likewise, Polikarpova, Furia, Pei, Wei, and Meyer found that "programmers are willing to write specifications if it brings tangible benefits to their usual development activities" [84]. For languages with first-class specifications (e.g. JML [32, 33, 50, 59, 85]) prior work has considered both static verification [11, 12, 22, 31, 40, 52, 57, 61, 68, 86] and random testing [16, 18, 24, 92, 101, 104]. An important question often raised here is: *why test when you could statically verify?* Of course, there are many reasons why one should still test. The ninth commandment of formal methods is "Thou shalt test, test and test again" [17]. Likewise, static verification tools remain difficult to use in practice [11, 14, 19, 25, 67]. Groce, Havelund, Holzmann, Joshi, and Xu found that, despite the considerable resources of NASA's Jet Propulsion Laboratory, it was "not feasible to produce a rigorous formal proof of correctness"" for mission-critical flight software and, instead, successfully employed random testing (amongst other techniques) [44]. Chamarthi, Dillinger, Kaufmann, and Manolios observed that users "have a difficult time determining whether the theorem prover failed because the conjecture is not true or because the theorem prover cannot find a proof" [20]. Petiot, Kosmatov, Giorgetti, and Julliand concur, claiming that "achieving a fully successful proof in practice needs a lot of tedious work and manual analysis of proof failures" [82, 83]. Their tool (StaDy) for testing ACSL specifications "can be used at early stages of specification, even when formal verification has no chances to succeed yet". As such, we argue that testing remains a useful tool given: (1) the ease with which it can be applied; and, (2) the insights it can provide (e.g., through counter-example generation).

Whiley is a programming language with first-class support for software specifications that is designed to simplify verification [75, 76, 77, 78, 79, 80, 93, 96, 97, 98]. For example, arithmetic types in Whiley consist of unbounded integers and explicit support is provided for distinguishing between pure *functions* versus side-effecting *methods*. The ultimate aim is that all programs written in Whiley will be verified at compile-time to ensure their specifications hold which, for example, has obvious application in safety-critical systems [23, 75]. Comparable systems to Whiley include ESC/Java [40], Spec# [11], Dafny [60], Why3 [39], VeriFast [51], SPARK/Ada [10], Frama-C [55],

Viper [66], KeY [2, 13], and RESOLVE [46, 89] (amongst others). Whiley has, for example, been used for teaching large introductory classes in formal methods [76, 79]. In this context, specification-based testing may seem an unusual choice. However, it provides complementary benefits such as offering students a stepping-stone prior to static verification. For example, loop invariants can be omitted when testing and, likewise, specifications can be weaker than required for static verification.

In this paper, we are interested in better understanding the trade-offs involved with automated testing compared with static verification. In particular, it seems reasonable to conclude that automated testing is less likely to find bugs than static verification. *But, how much less likely?* To explore this question, we consider automated specification-based testing for Whiley. This language is interesting for several reasons: firstly, it supports first-class specifications, which simplifies automated testing by providing an obvious test oracle [24]; secondly, the language was designed with static verification in mind and supports various features necessary for this (e.g. loop invariants); finally, a reasonable number of benchmarks written in Whiley are available for empirical study. This paper makes the following contributions:

- We present an open-source tool, QuickCheck for Whiley, for specification-based automated testing of Whiley programs which follows in the lineage of QuickCheck and related tools.[1] This exploits preconditions to determine valid inputs, and employs postconditions to check outputs for validity. A key usability concern is the ability to parameterise the input space in order to limit work done. To this end, we present novel approaches for parameterising the input spaces for references and lambdas, and introduce the concepts of *aliasing width* and *rotation*.
- We report on several large experiments evaluating the performance / precision trade-off across three benchmark suites. Our findings indicate automated testing is effective at bug finding, and that sampling offers useful performance gains for modest reductions in bug-finding capability. A number of actual bugs were also discovered in existing programs which further highlights the tools' value.

Finally, a curated artefact is available with which one can reproduce all the results from the experiments discussed in this paper.[2]

## 2 Background

The Whiley programming language was initially developed to facilitate static verification [76]. The Whiley Compiler (WyC) attempts to ensure that every function in a program meets its specification. When it succeeds in this endeavour, we know: 1) all function postconditions are met (assuming their preconditions held on entry); 2) all invocations meet their respective function's pre-condition; 3) runtime errors such as divide-by-zero, out-of-bounds accesses and null-pointer dereferences are impossible. Note, such programs may still loop indefinitely and/or exhaust available RAM.

---

[1] Now integrated with the Whiley compiler: http://github.com/Whiley/WhileyCompiler/
[2] http://ecs.victoria.ac.nz/~djp/files/QuickCheckForWhiley-2020-09-27.zip

**Finding Bugs with Specification-Based Testing is Easy!**

Given its original focus, Whiley supports certain features specifically to enable static verification. Whilst such features are necessary in this context, they are not always necessary in other contexts (such as automated testing). For example, one can ignore such features during automated testing and, hence, this offers a stepping stone prior to full static verification. In what follows, we draw specific attention to the manner in which such features manifest themselves.

**Pre- and Postconditions** Whiley supports *pre-* and *post-conditions* as follows:

```
1  function decrement(int x) -> (int y)
2  // Parameter must be positive
3  requires x > 0
4  // Return cannot be negative
5  ensures y >= 0:
6     return x - 1
```

Here, decrement() includes requires and ensures clauses which correspond (respectively) to its *precondition* and *postcondition*. The return value, y, may be used only within the ensures clause. The Whiley compiler statically verifies this function meets its specification (recall, integers are unbounded and cannot underflow).

The Whiley compiler reasons about functions by exploring their control-flow paths. As it learns more about their variables, it takes this into account. For example:

```
1  function max(int x, int y) -> (int z)
2  // Must return either x or y
3  ensures x == z || y == z
4  // Return must be as large as x and y
5  ensures x <= z && y <= z:
6     if x > y:
7        return x
8     else:
9        return y
```

Here, multiple ensures clauses are given which are conjoined to form the function's postcondition. We find that allowing multiple ensures clauses helps readability, and note that JML [34], Spec# [11] and Dafny [60] (amongst others) also permit this. Furthermore, multiple requires clauses are permitted in the same manner. As an aside, we note that the body of the function max() above is almost completely determined by its specification. However, in general, this it not usually the case and typically there is scope for significant variation between implementations.

**Type Invariants** Type invariants over data can be explicitly defined in Whiley:

```
1  type nat is (int n) where n >= 0
2  type pos is (int p) where p > 0
```

Here, the type declaration includes a where clause constraining the permitted values. The declared variable (e. g., n or p) represents an arbitrary value of the given type. Thus, nat defines the type of natural numbers. Likewise, pos gives the type of positive integers. Constrained types are helpful for ensuring specifications remain as concise as possible. For example, we can restate decrement() as follows:

```
1  function decrement(pos x) -> (nat n):
2     return x - 1
```

Observe that, in this example, there is no need to cast x to nat, as the compiler automatically determines that pos is a subtype of nat.

**Loop Invariants**   Whiley supports *loop invariants* expressed using one or more where clauses for statically verifying properties about loops. The following illustrates:

```
1  function sum(int[] xs) -> (nat r)
2  // All items in xs must be greater-or-equal to zero
3  requires all { i in 0..|xs| | xs[i] >= 0 }:
4     int s = 0
5     int i = 0
6     while i < |xs| where s >= 0 && i >= 0:
7        s = s + xs[i]
8        i = i + 1
9     return s
```

Here, a bounded quantifier enforces that sum() accepts only arrays of natural numbers.[3] A key constraint is that summing an array of natural numbers yields a natural number. The loop invariant s >= 0 && i >= 0 is provided to illustrate the syntax. The Whiley compiler statically verifies that sum() does indeed meet its specification. However, this *requires* the loop invariant given above for it to generate a sufficiently powerful verification condition for the proof. Unfortunately, writing loop invariants has long been recognised as challenging for both novices and experts alike [4, 11, 41]. However, loop invariants need not be given when automated testing is used. This is because testing operates with concrete values and, hence, loops are simply executed as normal. Indeed, when developing a function involving a loop (such as sum()), one can begin by omitting loop invariants and employ automated testing to help refine the function's specification. Loop invariants can then be developed once this is complete, giving an incremental route to static verification.

**(Uninterpreted) Functions**   An interesting aspect of static verification in Whiley (and related languages like Dafny) is that functions are treated as *uninterpreted* [45]. As for type checking this means that, when verifying code involving the invocation of some function, the verifier does not consider that function's implementation. Rather, it considers only the function's given specification and this is necessary to ensure verification is *modular* (i.e. intraprocedural). Amongst other things, this allows a function's implementation to change arbitrarily without affecting callers *provided it still meets its specification*. Of course, this requires the function to be fully specified which may not always be the case during development (such as when several functions are developed in tandem). In contrast, automated testing is more relaxed as, instead, it can simply execute the implementation as given. For example, consider the following:

---

[3] This precondition could equally have been expressed as type nat[].

**Finding Bugs with Specification-Based Testing is Easy!**

```
1  function max(int x, int y) -> (int r):
2    if x >= y:
3      return x
4    else:
5      return y
```

Whilst the above function is implemented correctly, it is missing a suitable specification. Perhaps this has arisen because it is, in fact, part of a larger function the developer is working on:

```
1  function max(int[] items, int i) -> (int r)
2  // Atleast one item must remain
3  requires 0 <= i && i < |items|
4  // Return greater than all remaining items
5  ensures all { k in i .. |items| | items[k] <= r }:
6    //
7    if (i+1) == |items|:
8      return items[i]
9    else:
10     return max(items[i], max(items,i+1))
```

As expected, max(int[],int) cannot be statically verified (yet) because the specification for max(int,int) (or lack thereof) yields insufficient information at the call site. Nevertheless the implementation of max(int,int) is correct and, despite the missing specification, one can employ automated testing to get some initial confidence in the correctness of max(int[],int). Again, this illustrates how automated testing provides a path towards static verification as specifications do not need to be fully fleshed out.

**Framing** A related aspect of static verification is the need for clarity around side-effects and framing [8, 53, 69, 74, 90]. Consider the following example (which also illustrates references in Whiley):

```
1  method copy(&int p, &int q):
2    *p = *q
```

This implements a very simple *method* which copies the value referred to by q over that referred to by p. We note that, whilst a function must be pure in Whiley, a method may have side-effects. As before, our interest lies in what this method *could do* versus *what it does*. To understand this, consider the following snippet:

```
1  ...
2  int x = 2
3  int y = 123
4  copy(&x,&y)
5  // Check expected outcome
6  assert (x == 123) && (y == 123)
```

Whilst the final assert holds, we cannot statically verify this (yet) as the specification for copy(&int,&int) remains insufficient. In such case, the static verifier makes worst-case assumptions at the call site regarding possible side-effects and, as a result, cannot conclude y is unchanged. In essence, we wish to say that &y is an *immutable* reference. Although Whiley has no support for expressing this at the time of writing, similar systems (such as Dafny) do.[4] The point is that, again, automated testing can proceed in the presence of incomplete specifications because it operates on the implementations as given.

## 3 Overview

We now present our tool, QuickCheck for Whiley (WyQC). Programs are first compiled into the Whiley Intermediate Language (WyIL) on which WyQC operates. Using this the tool discovers functions and methods which are then tested in order of appearance. When testing a given function or method, a *type generator* is constructed for each parameter which uniformly samples inputs at a given *sampling rate* (i.e. number of samples over domain size where a rate of 1.0 implies *exhaustive* testing). The tool then filters generated inputs according to the function or method's precondition. Thus, inputs which do not satisfy the precondition are discarded (so-called *meaningless* inputs [24]). Remaining inputs are then used to repeatedly execute the function or method using the interpreter. Any faults generated during this process (e.g. from out-of-bounds errors, division-by-zero, failed type invariants, failed assertions, etc) lead to test failures. All generated outputs are then checked against their postconditions and, again, test failures are raised accordingly. To illustrate, consider the following example adapted from the standard library:

```
1  function slice(int[] items, uint start, uint end) -> (int[] r)
2  // Given region to slice must make sense
3  requires start <= end && end <= |items|
4  // Size of slice determined by difference between start and end
5  ensures |r| == (end - start)
6  // Items returned in slice match those in region from start
7  ensures all { i in 0..|r| | items[i+start] == r[i] }:
8     ...
```

This illustrates several aspects of our approach. Firstly, generators for types int[] and uint are instantiated with the former producing inputs such as [], [0], [1], [0,0], and the latter producing inputs such as 0,1,2,3, etc. Observe that the exact range of inputs generated depends upon the given parameterisation of the input space (more on this later). The cross-product of generated inputs for each parameter is then filtered through the precondition. In this case, inputs such as [], 0, 0 and [1], 0, 1 pass through, whilst others such as [], 1, 0 and [1], 0, 2 are discarded because they do not meet the precondition.

### 3.1 Type Generation

Automatically generating parameter inputs is, in some cases, straightforward—e.g., the domain of values inhabited by the type bool is simply true and false. However, other data types describe infinite domains which, following Duregård, Jansson, and Wang, we map into *finite* domains via *parameterisation* [27, 37]. For example, the int type might map to the bounded domain -2 ... 2 (where the bounds are determined by the *upper* and *lower* parameters). Mapping types to finite domains simplifies input generation in several ways. Firstly, it is straightforward to enumerate all elements of the domain. Secondly, it is easy to sample uniformly from a finite domain and, perhaps more importantly, this can be done *without* enumerating the entire domain (e.g., using Knuth's Algorithm S [56]). However, whilst appropriate parameters are obvious in many cases, for others finding a sensible parameterisation is more challenging. We now consider the main type generators used, whilst noting various characteristics.

---

[4] In Dafny, reads / modifies clauses are used to signal what items a method may modify.

**Primitives**   These are straightforward: null maps to the singleton domain containing only the value null; bool maps to the domain of values true and false; and, as discussed above, int maps to a finite domain parameterised by an *upper* and *lower* bound. To give the user some ability to control the testing process, these are exposed as configuration parameters (which default to 3 and −3 respectively).

**Arrays**   Arrays are likewise relatively straightforward but, at a minimum, must be parameterised on their *maximum length* to yield finite domains. To generate inputs for an array type, we enumerate lengths up to the given bound and, for each, include the cross-product of inputs for each element. For example, assuming a maximum length 3, the type bool[] maps to: [], [true], [false], [true,false], ..., [false,false,false]. Again, the maximum length is a configuration parameter (which defaults to three). We note one characteristic is that this domain tends to be dominated by larger arrays. Thus, when sampling with a rate < 1.0 (i.e. not exhaustively examining the entire domain), there is a tendency for smaller arrays (esp. []) to be missed.

**Records**   Whiley supports both *open* and *closed* records. Closed records have a fixed set of fields whilst open records have a variable number. For example, {bool tag,int data} is a closed record. As for arrays, the generator for a closed record maps to the cross-product of its fields' domains. For example, {bool tag,int data}, maps to the values: {tag:false,data:-2}, {tag:true,data:-2}, ..., {tag:true,data:2}[5]

In contrast, {int x, int y, ...} is an open record whose valid instances must have *at least* integer fields x and y. Thus, both {x:0,y:0} and {x:0,y:0,z:0} are valid instances whilst {x:0} is not. Open records present a challenge because they permit an unbounded number of *field labels* which our tool cannot possibly generate. Fortunately, some relatively simple approaches exist: first, every open record has a set of fields explicit in the type from which a minimal instance can always be constructed (e.g. {x:0,y:0} for {int x, int y, ...}); secondly, to generate more complex instances, one could scan the enclosing source file and identify all field labels being used and generate instances using subsets from this. For simplicity, our tool adopts the first approach.

---

[5] Structural typing in Whiley offers some benefit here. In nominally typed languages like Java, generating object instances is limited by one's (in)ability to *construct* instances of a given class [72, 73]. For example, when a field is **private** and only a default constructor is provided, one may need to invoke sequences of methods to manipulate the field's value.

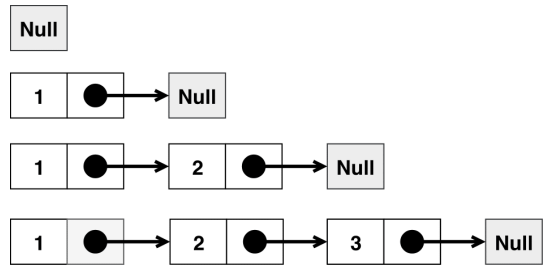

**Figure 1** Example values generated for the recursive List type.

**Unions** A union type is the composition of two or more types. For example, the type bool|int can hold either a boolean or an integer value. Again, it is relatively easy to generate an appropriate mapping from the union of element mappings. For example, bool|int can be mapped to: true,false,-3, …, 3. As before, one characteristic of this generator is that, when sampling with a rate $< 1.0$, it tends to favour larger domain(s). For example, when sampling from bool|int, it becomes more likely that *all* values from bool are missed with larger sizes (i.e. parameterisations) of int.

**Constrained Types** As discussed in section 2, Whiley supports user-defined types with invariants. Values for such types are first obtained using the generator for the *underlying* type and then filtered using the type invariant. The resulting values are then cached so as to avoid recomputing them needlessly again in the future. An important characteristic of this generator arises when sampling with a rate $< 1.0$. Since we cannot easily sample from the domain directly, we first sample the underlying domain and then filter as before. Unfortunately, this can result in relatively few inputs being found, depending on the relative sparsity of matching values in the underlying domain.

**Recursive Types** Whiley supports recursive structural types for implementing linked data structures, such as trees, etc. The following illustrates the syntax:

```
1  type List is null | {int value, List next}
```

Here, a List is either the null terminator or a record comprising the recursive structure {int value, List next}. Figure 1 illustrates some values for this type. Since recursive types can have infinitely many instances, we must find a suitable parameterisation for them. To do this, our tool bounds their maximum *depth* and exposes it as a configuration parameter (defaulting to three levels). Again, this domain tends to favour larger structures when sampling with a rate $< 1.0$. In particular, for binary or n-ary trees the domain is quickly dominated by larger structures as the depth increases.

**Reference Types** References refer to values allocated in the heap. From our perspective, there are some challenges in finding a suitable parameterisation. For example, consider the type &bool. *Does this represent a finite or infinite domain?* Technically speaking, it represents an infinite domain as there are infinitely many heap locations to which it could refer. Fortunately, in practice, there are only a finite number of meaningful locations where the exact number is determined by the function's signature overall. For example, consider the following function for swapping two locations:

```
method swap(&bool x, &bool y):
  bool temp = *x
  *x = *y
  *y = temp
```

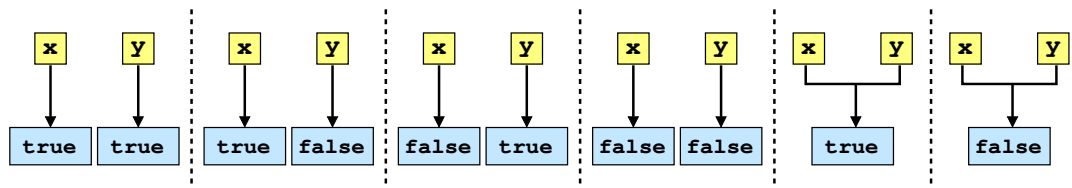

**Figure 2** Input combinations when calling swap for x and y.

Figure 2 shows all six possible input combinations where x and y can point to different cells or the same cell. Here, we have the four possible cases produced from the cross product of true and false, *along with two additional cases for aliasing*. In fact, for two references we have $2^2 + 2^1$ cases, for three we have $2^3 + 2^2 + 2^1$ cases, etc. We generalise this to $n^m + n^{m-1} + \ldots + n^1$ where $n$ is the size of the referent domain, and $m$ the number of reference parameters (of compatible type). Since this domain is already finite, further parameterisation is not strictly required. At the same time, it is an expensive domain and the ability to parameterise it can be useful (e.g. to reduce costs when exhaustive testing). As such, we parameterise the domain by bounding $m$ to a maximum value referred to as the *aliasing width*. In essence, for parameters of type &T this limits the number of distinct objects created per instance of T.

**Lambda Types** Whiley supports lambdas which can be passed as arguments. This presents yet another challenge, as synthesing arbitrary functions (even in simple cases) is intractable. To address this, we adopt a simple approach to parameterised generation of lambdas. Specifically, we map each value of the input domain to *exactly one* value in the output domain. To understand this, consider generating an instance of the type function(int)->(bool). Assuming integers between -1 and 1, there are three inputs and we can map them linearly like so: $\{-1 \mapsto \mathtt{false}, 0 \mapsto \mathtt{true}, 1 \mapsto \mathtt{false}\}$. Of course, for each lambda type, we would like to generate more than one valid instance! Therefore, we parameterize our approach using a *maximum rotation*, $k$, exposed as a configuration parameter. Specifically, we map the $i^{\text{th}}$ value of the input domain to the $i^{\text{th}} + k$ value of the output domain (modulo its size). Thus, we can generate arbitrarily many instances of a given lambda type and the maximum rotation (which defaults to 2) simply limits the maximum number of instances to generate.[6] Finally we note that, at this time, lambdas in Whiley do not support explicit pre-/post-conditions.

---

[6] Another approach would be to search the given compilation unit for matching instances of the lambda. Unfortunately, the challenge here is ensuring compatible specifications.

■ **Table 1** Descriptions of the Whiley benchmark suite.

| Name | Description | LOC |
|---:|---|---:|
| 001_average | Average over integer array | 43 |
| 002_fib | Recursive Fibonacci generator | 17 |
| 003_gcd | Classical GCD algorithm | 40 |
| 004_matrix | Straightforward matrix multiplication | 139 |
| 006_queens | Classical N-Queens problem | 54 |
| 007_regex | Regular expression matching | 80 |
| 008_scc | Tarjan's algorithm for finding strongly connected components | 201 |
| 009_lz77 | LZ77 compression / decompression | 201 |
| 010_sort | Merge Sort | 102 |
| 011_codejam | Solution for Google CodeJam problem | 118 |
| 012_cyclic | Cyclic buffer | 165 |
| 013_btree | Binary search tree with insertion / lookup | 172 |
| 014_lights | Traffic lights sequence generator | 64 |
| 015_cashtill | Simple change determination algorithm | 219 |
| 016_date | Gregorian dates | 88 |
| 017_math | Simple math algorithms | 193 |
| 018_heap | Binary heap data structure | 168 |
| 022_cars | Controlling cars on bridge problem | 54 |
| 023_microwave | Classical microwave state machine | 87 |
| 024_bits | Algorithms for bit arrays | 99 |
| 025_tries | String trie with lookup / insertion | 166 |
| 026_reverse | Reversing an array | 58 |
| 027_c_string | Model of C strings | 68 |
| 028_flag | Dutch national flag Problem | 99 |
| 029_bipmatch | Perfect matching for bipartite graphs. | 217 |
| 030_fractions | Big rationals | 59 |
| 032_arrlist | ArrayList implementation | 88 |
| 033_bank | Simple account system | 72 |
| 102_conway | Conway's Game of Life | 168 |
| 104_tictactoe | Tic-Tac-Toe | 111 |
| 107_minesweeper | Minesweeper | 100 |

## 4 Evaluation

We have evaluated our tool using two existing suites of Whiley programs. Our evaluation explores the effectiveness of the tool in finding bugs, and we report on several real bugs found using it. A curated artefact is also available with which one can reproduce all the results of this evaluation (depending on available time).[7] The test suites used for our evaluation are:

---

[7] http://ecs.victoria.ac.nz/~djp/files/QuickCheckForWhiley-2020-09-27.zip

**Finding Bugs with Specification-Based Testing is Easy!**

▮ **Table 2** Summary performance and precision results for the compiler fail tests with different scopes and sampling rates. *Time* reports a sum of the mean execution time (in seconds) for each test across 10 actual runs with 5 warmup runs discarded and a timeout of 120 s for all 15 runs. *Precision* indicates the percentage of tests correctly identified as erroneous by WyQC.

|           | tiny  | small  | medium |       | large  |        |        | huge   |        |        |        |
|-----------|-------|--------|--------|-------|--------|--------|--------|--------|--------|--------|--------|
|           | 1.0   | 1.0    | 0.1    | 1.0   | 0.01   | 0.1    | 1.0    | 0.001  | 0.01   | 0.1    | 1.0    |
| time      | 0.0 s | 0.03 s | 1.0 s  | 8.5 s | 16.3 s | 33.1 s | 37.4 s | 14.7 s | 47.9 s | 56.4 s | 95.4 s |
| precision | 69 %  | 81 %   | 83 %   | 83 %  | 82 %   | 82 %   | 83 %   | 82 %   | 83 %   | 83 %   | 84 %   |

**Compiler Tests** The Whiley compiler ships with 1000+ test cases, where each test is a Whiley program which either *compiles* without problem, or *fails* with one or more errors. The latter (referred to as *fail tests*) are used to ensure the compiler correctly catches different types of error. From our perspective, these are of more interest because they can provide insights into the precision of automated testing (compared with static verification). From the full test suite, there are 338 of these so-called fail tests. Many of these fail tests check for errors which arise early in the compilation pipeline (e.g. parsing, type checking, etc). However, 126 / 338 require static verification to identify the errors they contain and it is these on which we focus. In particular, by replacing static verification with WyQC we can evaluate its precision (i.e. ability to uncover these errors).

**Benchmark Suite** The Whiley Benchmark Suite consists of 31 small benchmarks totalling over 4KLOC of code [100]. These cover various problems including: *N-Queens*, *LZ77 compression*, *matrix multiplication*, *Conway's Game of Life*, *tic tac toe*, *merge sort*, etc. Table 1 provides more details of the benchmark programs tested. We note that some of these programs have not been statically verified using Whiley (e.g., because they currently take too long to complete).

**Mutants Suite** Using the above benchmark suite, we automatically generated a set of 1826 *mutants* by applying a range of mutations to the original benchmarks. The mutations considered were relatively simplistic, and included swapping operators (e.g. != for ==, < for >=, && for ||, etc), altering integer and boolean constants, swapping quantifiers (e.g., some for all), etc. For each benchmark, the domain of all possible single mutations was constructed. These were then uniformly sampled to ensure at most 100 mutants per benchmark, giving 1826 overall. Unfortunately, one cannot be certain whether a given mutant will actually be incorrect (in some sense) or not. For example, a mutation to a function's implementation may still (by coincidence) meet its specification and, hence, it is hard to judge how many could realistically be considered as erroneous (more on this later).

**Methodology** For each test suite, we performed experiments using different scopes and sampling rates for input generations. Five scopes were considered:

|        | Int Bounds         | Array Lengths    | Type Depths      | Alias Width      | Max Rotation     |
|--------|--------------------|------------------|------------------|------------------|------------------|
| tiny   | $\langle 0\ldots 0\rangle$  | $\langle 0\ldots 0\rangle$ | $\langle 0\ldots 0\rangle$ | $\langle 0\ldots 0\rangle$ | $\langle 0\ldots 0\rangle$ |
| small  | $\langle -1\ldots 1\rangle$ | $\langle 0\ldots 1\rangle$ | $\langle 0\ldots 1\rangle$ | $\langle 0\ldots 1\rangle$ | $\langle 0\ldots 1\rangle$ |
| medium | $\langle -2\ldots 2\rangle$ | $\langle 0\ldots 2\rangle$ | $\langle 0\ldots 2\rangle$ | $\langle 0\ldots 2\rangle$ | $\langle 0\ldots 2\rangle$ |
| large  | $\langle -3\ldots 3\rangle$ | $\langle 0\ldots 3\rangle$ | $\langle 0\ldots 3\rangle$ | $\langle 0\ldots 3\rangle$ | $\langle 0\ldots 3\rangle$ |
| huge   | $\langle -4\ldots 4\rangle$ | $\langle 0\ldots 4\rangle$ | $\langle 0\ldots 4\rangle$ | $\langle 0\ldots 4\rangle$ | $\langle 0\ldots 4\rangle$ |

For example, in the large scope, integers in the range -3...3 are considered as are arrays with lengths between 0...3 and recursive types with depths between 0...3, etc. In addition, various sampling rates were employed at each scope (e.g. 0.001, 0.01, 0.1 and 1.0 for the *huge* scope). As outlined briefly in section 3, this determines how many inputs are actually selected from the input domain. For example, at a sampling rate of 1.0 *every input* is selected, whilst at a rate of 0.1 only *one in every ten inputs* is selected (uniformly at random). As such, small sampling rates were only considered at larger scopes where sufficient values would be generated. Finally, performance data was averaged over ten executions, with five warm up runs discarded beforehand. This was done to ensure steady state performance was reported and to ensure low variance [43]. The experimental machine was a Dell OptiPlex 7050 with eight Intel i7-7700 x86-64 cores at 3.6 GHz with 8 GB SRAM running Arch Linux.

### 4.1 Experiment I — Compiler Fail Tests

The purpose of this experiment was to compare automated testing with WyQC against the existing static verifier. In particular, to determine the proportion of fail tests which WyQC correctly identifies as erroneous for different sampling rates. The following illustrates a fail test which WyQC correctly identified as erroneous:

```
1  function f(int x) -> (int r) ensures r > 1:
2      return x
```

This is erroneous as its postcondition requires r > 1, but has no precondition enforcing x > 1. Thus, if x==0 its precondition holds but its postcondition fails, etc.

Table 2 reports the results. Note, timeouts have a dampening effect for larger scopes (i.e. because such tests could run for significantly longer). Also, WyQC is reported as correctly identifying a fail test only if it did so for all 10 actual test runs. Looking at table 2, there are a few interesting observations. As expected, the *tiny* scope doesn't perform well. But, perhaps surprisingly, the other scopes are fairly comparable in terms of precision, with the medium scope offering a good balance. In fact, the precision results are better than it first appears. Specifically, we examined by hand those tests that WyQC could not identify as invalid and found that most ($\frac{17}{21}$) could *never* be caught no matter how large the scope. Listing 1 illustrates such an example. To better understand this, consider the specification of f() in the figure. This does not ensure the return value r is within bounds of xs (i.e. since r==|xs| is permitted). When this

**Finding Bugs with Specification-Based Testing is Easy!**

■ **Listing 1** Illustrating an invalid test for which WyQC could never find a failing input due to the particular implementation of f(). For example, if f() returned 1 instead (which is permitted under its specification) then the problem in g() would be detected immediately.

```
1  function f(int[] xs) -> (int r) requires |xs| > 0
2  ensures r >= 0 && r <= |xs|:
3      return 0
4
5  function g(int[] xs) -> (int r) requires |xs| > 0:
6      int[] indices = [f(xs)]
7      return xs[indices[0]]
```

value is subsequently used in g() to index xs, a static verifier must conclude it could be out-of-bounds and, hence, report an error (recall the discussion on uninterpreted functions from section 2 for more on this). For the given implementations of f() and g() this cannot (by coincidence) arise at runtime and, hence, cannot be detected by WyQC.[8]

Another common class of examples which could never be caught by WyQC arise from missing loop invariants. The following illustrates:

```
1  function count(nat n) -> (int r) ensures r == n:
2      nat i = 0
3      while i < n:
4          i = i + 1
5      return i
```

This function will not statically verify because it is missing a required loop invariant (i.e. where i <= n) and, for this reason, is considered erroneous (recall from discussion of loop invariants in section 2 that they are needed specifically for static verification). However, there is no actual bug! Thus, automated testing could not find a problem no matter how large the scope.

**4.2 Experiment II — Benchmark Suite**

The purpose of this experiment was to evaluate the performance of WyQC for differing scopes and sampling rates across a range of more realistic benchmark programs. In this experiment, the main() method for each benchmark was skipped as this often forced a timeout (e.g. for N-Queens it forces WyQC to try and solve the problem for a board size of 10). The results are reported in figure 3 (with raw data as table 3 in the Appendix). They indicate that, for many benchmarks, exploring beyond the *medium* scope is often expensive. Observe missing columns at larger scopes (see e.g. 004_matrix, 006_queens, 007_regex, 008_scc, etc) indicate timeouts occurred (i.e. that automated testing was intractable in the given timeframe). Taking 008_scc as an illustrative example, WyQC can actually check this benchmark for the medium scope at a sampling rate of 0.1 *but*

---

[8] In principle, an automated testing tool could attempt to catch such problems by fuzzing against specifications as well. That is, instead of executing the given function, one could automatically generate arbitrary return values which meet its specification.

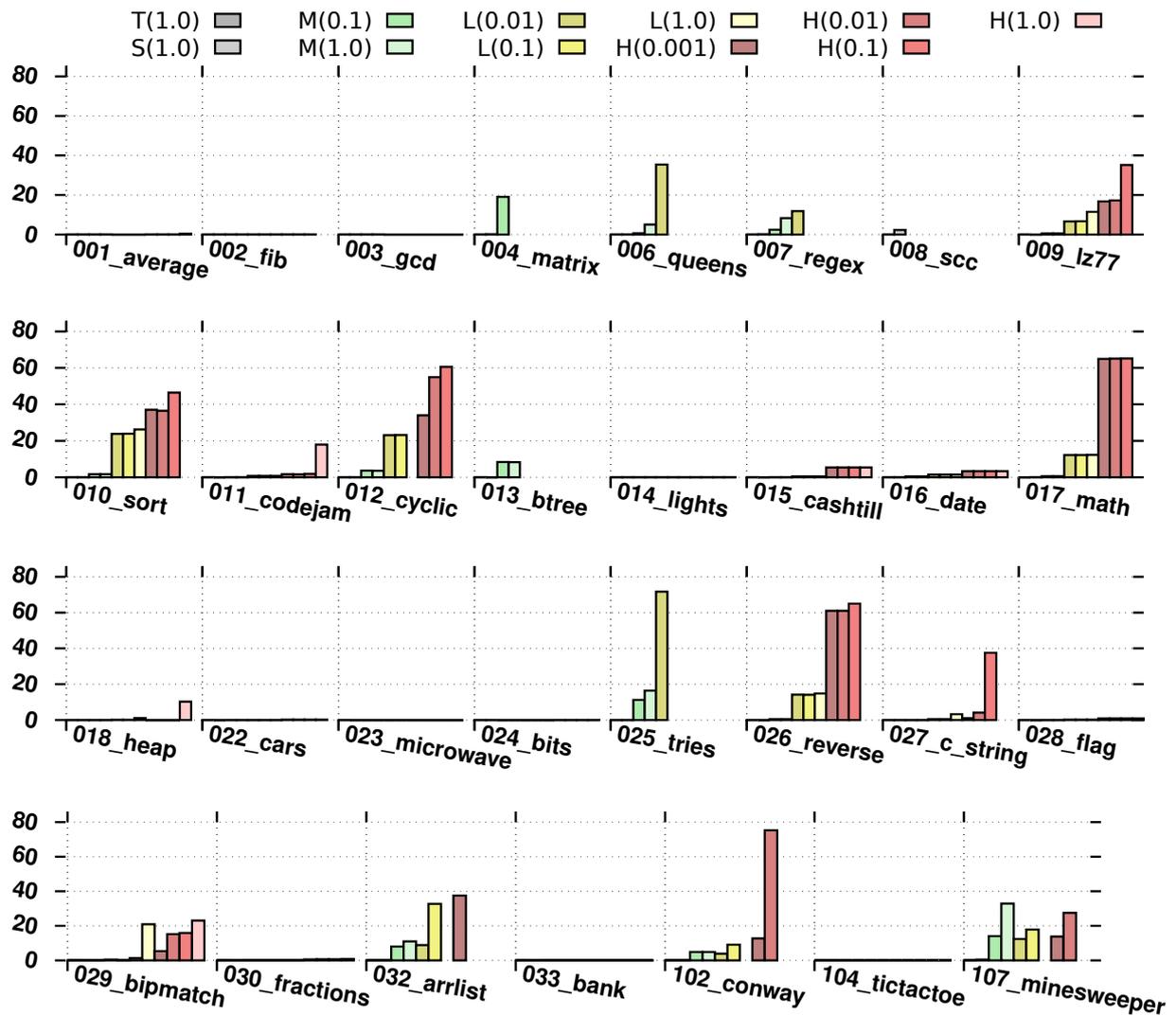

**Figure 3** Performance results for benchmarks with differing scopes and sampling rates (T=Tiny, M=Medium, L=Large, H=Huge). Mean execution time (in seconds) is reported across 10 actual runs with 5 warmup runs discarded. A timeout of 1200 s was employed for the 15 runs of each benchmark and sampling / scope configuration. Missing values indicate timeouts occurred.

*requires around 43 min to do so*. The challenge stems from the following data structure used during the depth-first traversal in `008_scc`:

```
1  type State is { Digraph graph, bool[] visited,
2    bool[] inComponent, int[] rindex, Vector<nat> stack,
3    int index, int cindex
4  } where |visited| == |graph| && |inComponent| == |graph|
5    where |rindex| == |graph|
6    where all {k in 0..stack.length | stack.items[k] < |graph|}
```

# Finding Bugs with Specification-Based Testing is Easy!

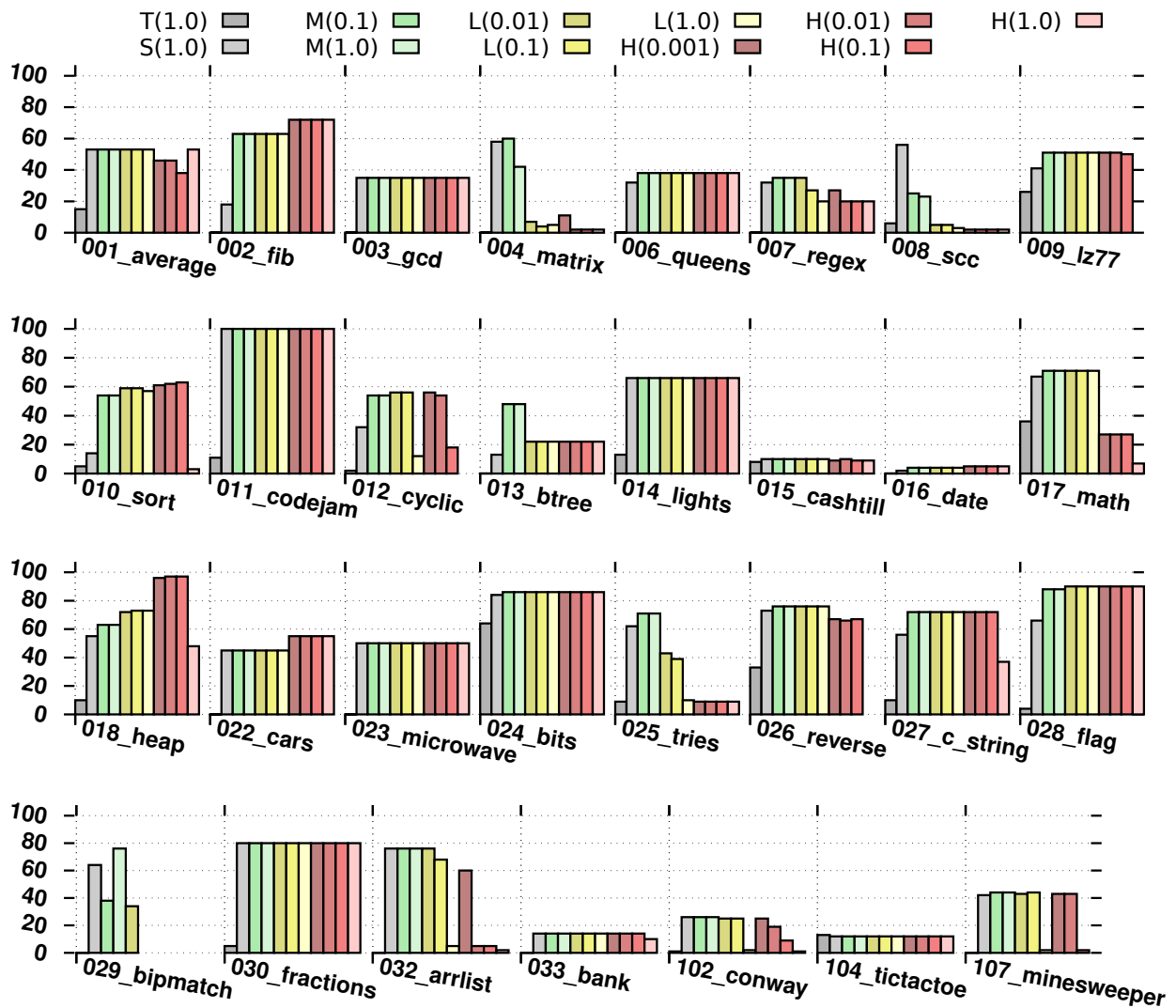

**Figure 4** Illustrating, for each benchmark and scope/sampling configuration, percentage of mutants identified as erroneous (T=Tiny, M=Medium, L=Large, H=Huge).

This type illustrates a number of interesting points. Most importantly, the presence of four arrays (since `Vector<nat>` internally contains an array) corresponds to a very large space from which to draw input values. As such, low sampling rates (e.g. < 0.001) are effective at sampling from this domain in reasonable time. However, finding input values which meet the given type invariants is challenging and, at low sampling rates, relatively few are typically encountered. The difficulty arises from the `State` requirement that three arrays have the same size where, for increasingly larger scopes, the probability of meeting this requirement decreases. Note, this data structure can be restructured to avoid this entirely by replacing `visited`, `inComponent` and `rIndex` with a single array `{bool visited,bool inComponent,int rIndex}[]`.

### 4.3 Experiment III — Mutants Suite

The purpose of this experiment was to evaluate the bug-finding ability of WyQC for differing scopes/sampling rates across a range of realistic benchmark programs. In this experiment, only a single run was used for each mutant and scope/sampling configuration, with a timeout of 60 s employed. The results are reported in figure 4 (with raw data as table 4 in the Appendix) and there are several points to make:

**(Timeouts)** The effect of timeouts is important to understand here. Firstly, we cannot avoid timeouts because some mutants enter infinite loops. Secondly, we considered a timeout to indicate that WyQC failed to identify a mutant as invalid. This has a "lowering" effect on precision for larger scopes. In particular, larger scopes typically have more timeouts and, hence, lower precision. For example, looking at `007_regex`, the best precision was not obtained with the *huge* scope, but the *medium* or *large* ones.

**(Precision)** The maximum achievable precision for WyQC across this mutant suite remains unclear. This is because some generated mutants remain correct (with respect to their specification), but with slightly altered behaviour. As such, we cannot tell, for each benchmark, whether WyQC could ever achieve 100 % precision (see below on efforts to bound this).

**(Problems)** Some benchmarks have consistently low precision and this typically stems from some unmet requirement. For example, `104_tictactoe` requires arrays of size 9 to be considered before any valid inputs could be found. Likewise, `015_cashtill` requires arrays of size 8, etc.

Overall, the results indicate the tool finds bugs with high probability in many cases. We also observe a diminishing benefit from larger scopes with $\frac{23}{31}$ benchmarks achieving best precision with *medium* scope. Furthermore, it is clear that sampling offers a good trade-off in terms of performance and precision. Finally, in an effort to estimate an upper bound on precision, we manually inspected a random sample of $\frac{33}{326}$ mutants (≈10 %) for which the tool never identified a bug. Our findings indicated that only $\frac{9}{33}$ of these mutants (≈27.3 %) contained *detectable* bugs. For example, an undetectable change witnessed in `016_date` saw the constant `SEP` changing from `9` to `10`. Since the month is only ever referred to by this constant, this change is not detectable as it uniformly updated all specifications and implementations. In contrast, another mutant witnessed had `day <= 29` changed to `day < 29` which is detectable as a post-condition violation (but requires a sufficiently large integer).

### 4.4 Real-World Examples

As a result of running QuickCheck for Whiley for the evaluation, numerous bugs were discovered within the various test suites and also the standard library. Subsequently, these bugs were logged as issues in their relevant GitHub repositories along with counterexamples generated by the tool. Bug fixes were then created as pull requests and accepted after review.

**Finding Bugs with Specification-Based Testing is Easy!**

**Benchmarks** Examples of issues raised as a result of using the tool include: https://github.com/Whiley/WyBench (#23, #25, #28, #30, #32, #35, #37, #38, #40); https://github.com/Whiley/WhileyCompiler (#855, #857, #859).
For example, the original specification for `010_sort` was insufficient:

```
1  function sort(int[] items, int start, int end) -> (int[] rs)
2  ensures sorted(rs,start,end)
3  ...
```

The intention here is that all items between start and upto (but not including) end are sorted in the resulting array rs. However, WyQC quickly identified that the specification fails to enforce that `0 <= start && start <= end && end <= |items|` holds.

A similar issue was found by WyQC in the following specification given for `018_heap`:

```
1  type Heap is { int[] data, int len}
2  // Items on left branch are below their parent's item
3  where all {i in 0..len | (2*i)+1<len ==> data[i]>=data[(2*i)+1]}
4  // Items on right branch are below their parent's item
5  where all {i in 0..len | (2*i)+2<len ==> data[i]>=data[(2*i)+2]}
```

Again, the specification fails to enforce that `0 <= len && len <= |data|` which was easily spotted by WyQC.[9]

**Standard Library** The tool has also identified a number of bugs in the standard library: https://github.com/Whiley/STD.wy (#4, #10).[10] A key benefit with automated testing here was its ability to provide concrete counterexamples when filing bug reports. The following illustrates a function from the standard library for which a bug was identified using WyQC (github.com/Whiley/STD.wy/issues/10):

```
1   function set<T>(Vector<T> vec, int ith, T item) -> (Vector<T> res)
2   // Index must be within array bounds
3   requires ith >= 0 && ith < |vec.items|
4   // Length of vector unchanged
5   ensures vec.length == res.length
6   // All items below ith remain unchanged
7   ensures array::equals<T>(vec.items,res.items,0,ith)
8   // All items above ith remain unchanged
9   ensures array::equals<T>(vec.items,res.items,ith+1,res.length)
10  // Ith element assigned item
11  ensures res.items[ith] == item:
12     vec.items[ith] = item
13     return vec
```

Whilst the implementation of this function is fairly straightforward, it is not easy to see that there is a problem with its specification. The actual output generated by WyQC for this function is:

---

[9] Recall that some benchmarks have not been statically verified due to limitations with Whiley's verifier (hence why bugs were available to find).
[10] Again, finding bugs was possible as the standard library has yet to be statically verified.

```
1  ./src/whiley/std/array.whiley:39: negative array range
2  where all { i in start..end | lhs[i] == rhs[i] }
3                                 ^^^
4  Stack Trace:
5  --> std::array::equals([-1],[-1],1,0)
6  --> std::vector::set({items=[-1], length=0},0,-1)
```

Here, we see the tool reports a stack trace for the given violation which includes concrete parameter values. Using this information, the bug was easy to identify and fix. The issue is that we should have ith < vec.length rather than ith < |vec.items| above.

### 4.5 Limitations

Finally, we note some limitations of the given approach which warrant further consideration. For example, floating point types have not been considered simply because Whiley does not have them. Likewise, concurrency has been ignored because Whiley (at the time of writing) assumes sequential execution. If such support is eventually added, WyQC will need to be extended accordingly. Also, unlike other specification languages (e.g. Dafny), syntax for *framing* has yet to be included in Whiley (though this does not pose any specific challenges here) [8, 53, 69, 74, 90]. Finally, supporting native methods (i.e. those implemented externally by the system) does present some challenges, such as accounting for the effects of reading / writing files.

## 5  Related Work

**QuickCheck**   This tool was originally implemented by Koen Claessen and John Hughes for testing Haskell programs and has received widespread use in both academia and industry [28]. For example, it has been used to test Galois' Cryptol compiler [48], Ericsson's H.248 Media Proxy interface [6], and more [7, 29, 47]. Whilst QuickCheck for Haskell operates in a roughly similar fashion to that described here, there are some differences. In particular, one must provide user-defined properties (given as Haskell functions) to act as the test oracle which contrasts with Whiley, where pre- and post-conditions fulfil this role. QuickCheck for Haskell also generates random inputs for primitive types uniformly, but requires generators be provided for user-defined types. This allows fine-grained control over the distribution of test values. Indeed, the authors argue that it is "meaningless to talk about random testing without discussing the distribution of test data" [28].[11] In particular, since the presence of arbitrary constraints can skew the generation of values. To understand this, consider a requirement for an *ordered* list. Smaller lists will dominate the set of generated values because, as the length increases, the probability of generating an ordered list decreases. Indeed, lists which have at most one element are, by construction, ordered. In contrast, our approach is to generate values for all types, including user-defined types. Whilst this simplifies the process of testing, it would be useful to experiment with user-defined generators in the future.

---

[11] Though we argue this work and others [102, 104] demonstrates significant value can be obtained from automated testing without consideration of test value distribution.

The authors of QuickCheck chose specifically not to employ coverage metrics to keep the tool simple, and also because "their is no clear reason to believe that doing so would yield better results". They also comment that "pure functions are much easier to test than side-effecting ones" [28]. This concurs with our findings where methods are generally more awkward to handle. Indeed, this highlights a key advantage of Whiley over other imperative languages: first-class support for pure functions. Another important aspect of QuickCheck is the concept of *shrinking* failing test inputs to simplify debugging [47]. Whilst WyQC does not currently support shrinking, this would also be useful to include in the future.

Claessen, Smallbone, and Hughes subsequently developed a follow-on tool, QuickSpec, which attempts to infer properties for use with QuickCheck [30]. The motivation for this was that "coming up with formal specifications is difficult, especially for untrained programmers". The tool works only on pure functions and, through testing, partitions inputs into equivalences classes from which observed properties are extracted. Finally, Hughes, Norell, Smallbone, and Arts have recently explored further improvements [49]. QuickCheck has a tendency of fixating on one kind of bug over others — something exacerbated by test case reduction (i.e. shrinkage). To understand this, suppose a single call to some function is enough to trigger a bug. The probability a test sequence includes such a call increases with sequence length. Furthermore, test-case reduction tends to reduce a sequence to *just this call* even if an earlier part of the sequence causes a different bug. As such, their new approach generalises sequences into patterns which subsequent sequences cannot match. This yields better breadth of bug finding, though it does still mask some bugs. For example, if there are different kinds of bug which can be triggered by our function call (e.g. through differing parameters), this approach is still likely to only find one of them.

Finally, QuickCheck for Dafny was recently developed by Cameron and, given the obvious similarities between Dafny and Whiley, shares overlap with that presented here [19]. Cameron argues the "verification process can be slow and if a failure occurrs it can be unclear whether this was due to the code being wrong, the specification being wrong or that the prover is not powerful enough". That said, being developed as part of an honours thesis, the tool itself remains relatively simplistic. For example, it does not support references or lambdas (both of which are found in Dafny) and, for integers, enumerates $\{-1, 0, 1, 10\}$ and then chooses randomly from $\{-5000 \ldots 5000\}$.

**Java Modelling Language** Perhaps the first tool to exploit JML specifications for automated testing was JMLUnit [24]. Whilst somewhat similar to that presented here, the tool was fairly primitive by comparison. For example, the only integers considered by JMLUnit were $\{-1, 0, 1\}$, whilst the default set of strings was $\{\textbf{null}, ""\}$. Zimmerman and Nagmoti noted that JMLUnit suffers from "the need to manually write significant amounts of code to generate non-primitive test data objects" [104]. Their tool, JMLUnitNG, automatically generates non-primitive test inputs by recursively constructing objects as needed. However, cyclic dependencies and classes with no public constructors were a problem. Likewise, no attempt was made to ensure a fair spread of subclasses being instantiated. Nevertheless, their preliminary results over a small benchmark suite indicated it outperformed the original.

Korat is another interesting tool for specification-based testing of Java, and a prototype was developed for JML [18]. The tool focuses on integers, arrays and objects though doesn't consider lambdas or problems related to aliasing. A mechanism for pruning the input space was employed based on instrumenting variable accesses. Specifically, for a given test run, we know that unused variables need not be further enumerated giving sizeable reductions in the search space. Empirical data was also reported over some small benchmarks (e.g. a binary tree implementation) suggesting pruning to be effect effective. Bouquet, Dadeau, and Legeard focused on obtaining sufficient coverage of the specification [16]. Firstly, a form of *decision coverage* was employed to ensure inputs exercise all decisions in the precondition. Secondly, a *boundary analysis* was performed to guide generation of inputs. Unfortunately, the evaluation consisted of only a single benchmark and, hence, further empirical work is needed to explore the effectiveness here. Finally, other tools in a similar vein include: Jartege, which provides some simple statistical controls (e.g. weight) over object creation [71]; and, JMLAutoTest, which samples a few initial inputs in an effort to guess a good partitioning [101].

TestEra takes a similar approach to those based on JML, and employs Alloy for expressing specifications and its SAT-based model checker for input generation [54]. Unfortunately, Alloy's limited support for integers and arrays present significant issues, and the empirical results suggest Korat (above) is more effective. Yatoh, Sakamoto, Ishikawa, and Honiden argue a major drawback of many property-based testing tools is that they "require developers to write generator functions for user-defined types" [102]. Like Quickcheck, their tool automatically generates inputs for all types, though requires properties be provided to act as an oracle. Aichernig and Schumi applied property-based testing to web applications [3]. Earle and Fredlund applied QuickCheck for Erlang to Java programs using a library which reduced boilerplate [38]. Finally, Börding, Haltermann, Jakobs, and Wehrheim consider contracts expressing relations between two methods (e.g. equals() and hashCode()) and provide a DSL for expressing them along with a random testing tool [15].

**Other Tools**  AutoTest targets the Eiffel programming language which, like Whiley, supports first-class specifications that act as an oracle [62, 65]. Since Eiffel is object-oriented, AutoTest maintains an object pool from which instances of a given type are extracted (roughly following RANDOOP — see below). New objects are added with some given probability to diversify the pool, and are constructed by randomly choosing a constructor and recursively obtaining arguments. In contrast, values of primitive types are selected at random from preset domains (e.g. integer ranges, etc). A notion of *object distance* is used within AutoTest which characterises when two values are "close together" to ensure an even spread of values across the domain. Ciupa, Leitner, Oriol, and Meyer report on what is perhaps the largest experimental study of AutoTest [26]. Using a grid of 32 dual-core machines, eight production classes (e.g. from the standard library) totalling roughly 5KLOC were tested across a range of configurations for around 1500hrs of compute time. The results suggest that most bugs are found relatively quickly, and that the choice of random seed can have a noticeable impact.

Chamarthi, Dillinger, Kaufmann, and Manolios, Chamarthi, Dillinger, and Manolios focus on providing a better experience for static verification through the use of testing to find concrete counterexamples [20, 21]. However, their tool is not fully automated and, for anything other than primitive types, the user must provide custom generators. Petiot, Kosmatov, Giorgetti, and Julliand describe a tool, StaDY, for automated testing in the context of Frama-C—a verification framework for C [83]. This employs concolic testing, and operates on a subset of the ANSI C Specification Language (ACSL). They argue that testing complements verification by helping distinguish the different reasons why a proof fails [82]. For example, if testing finds a counterexample, that helps by providing a concrete test failure. At the same time, failing to find a counterexample also offers guidance as it suggests the program is correct but the specification too weak. Indeed, Beyer, Dangl, Lemberger, and Tautschnig take this idea one step further by generating actual test cases to give to developers [14]. They argue this helps bridge the gap between developers who are more comfortable debugging tests than interacting with verification tools.

RANDOOP applies feedback-directed random test generation to Java [72, 73]. Generating input values is challenging because the possible states of an object cannot necessarily be reached through construction alone. For example, the object may need to be constructed and then mutated using one or more methods. RANDOOP constructs objects iteratively by randomly selecting a method or constructor to invoke using previously computed objects as inputs. Unfortunately, RANDOOP does not have an obvious test oracle to rely upon. For example, whilst throwing a `NullPointerException` could be considered a test failure, this could equally be considered correct behaviour in Java. Instead, RANDOOP relies on user-defined *contracts* which express invariant properties that must hold before and after a method's execution. To simplify things, a number of built-in contracts are provided (e.g. checking `Object.hashCode()` does not throw an exception, etc). Finally, JCrasher is a similar tool which, again, generates random inputs based on a method's type signature and employs heuristics to decide what constitutes a bug. For example, throwing an `ArrayOutOfBoundsException` is considered a bug, whilst throwing an `IllegalArgumentException` indicates an invalid input (i.e. a precondition violation). Indeed, the authors note that "in the absence of explicit preconditions it is not generally clear whether an exception is a sign of programming error". An unusual aspect of JCrasher is that it employs estimation and planning to help ensure maximum benefit in a given timeframe.

**Input Generators** The underlying approach taken in QuickCheck for Whiley was inspired by the approach of Duregård, Jansson, and Wang which first features in the FEAT library [37]. This approach maps enumerable value types (such as integers, arrays, records, etc) to integer indices. From the perspective of automated testing, this offers several advantages as it enables efficient "random access" to the set of all values. For example, one can easily sample uniformly from the set of all values using classical algorithms such as (e.g., using Knuth's Algorithm S [56] or Waterman's *reservoir sampling* Algorithm R [94]). To understand this, consider the following language of integers and products:

$$v ::= v_1 \times v_2 \mid 0 \mid 1 \mid 2 \mid \ldots$$

To map this language to a finite set of integer values we limit the integers to some upper bound $n$, such that indicates $0 \leq k < n$ map directly to this bounded set. Then, indices $n \ldots n^2$ represent the set of all pairs of integer values, and so on. Duregård, Jansson, and Wang observe that one can index any finite (countable) structure and, for example, applied this to enumerating Abstract Syntax Trees.

The later work of Claessen, Duregård, and Palka took this general theme one step further by considering the efficient generation of inputs which meet some predicate [27]. This is certainly relevant to Whiley, where inputs must be generated which meet given preconditions and type invariants. The general approach based around aggressive pruning in a similar fashion to Korat [18]. For example, consider the following type:

```
1  type { int item, int[] items }
2  where item >= 0 && some { k in 0..|items| | items[k] == item }
```

To generate valid instances of this type one can begin by *generating* some number of values corresponding to the given type and then *filter* them using the invariant. But, this is reasonably inefficient. For example, suppose we generated {item:-1,items:[]} and {item:-1,items:[-1]}. The *generate-and-filter* approach must generate both before filtering them. However, having generated {item:-1,items:[]} one could easily conclude any input {item:-1,...} is invalid. This requires an awareness that only item was used in falsifying the predicate and, hence, the value for items is a *don't care*. A key downside to this approach is that uniform sampling becomes more tricky. The issue is that pruning like this prevents one from easily mapping values to integer indices (which is key to efficient sampling). Claessen, Duregård, and Palka employ a backtracking algorithm which retains reasonable efficiency at the cost of predictable uniformity of sampling.

Finally, we note the promising results of the Luck language for writing domain-specific generators which, over a small set of benchmark examples, was able to demonstrate competitive performance with hand-written (QuickCheck) generators [58].

# 6 Conclusion

We have considered the question of how effective automated specification-based testing is at finding bugs. To this end, we developed an automated testing tool for Whiley and have reported here on several large experiments investigating its effectiveness at finding bugs across difference scopes and sampling configurations. Our findings indicate automated testing can find bugs with high probability in many cases. We also found a diminishing benefit from larger scopes and that sampling offered a good trade-off in terms of performance and precision. Using our tool, we also discovered a number of actual bugs in existing Whiley programs, including the standard library.

We have also provided here a technical discussion of the tool's implementation. Whilst the tool is, in some sense, comparable with QuickCheck for Haskell there are some notable differences. In particular, since Whiley supports explicit pre- and post-conditions, this eliminates the need for user-defined properties to be supplied as a test oracle (i.e. as required for QuickCheck for Haskell). Indeed, QuickCheck for Whiley is a fully automated testing tool and does not require generators be supplied for user-defined types, and its implementation adopts novel approaches for dealing with references and lambdas. Finally, the tool has been released as part of the (open source) Whiley Compiler available at https://github.com/Whiley/WhileyCompiler/. In addition, an artefact has been prepared for reproducing the results of our evaluation.[12]

**Acknowledgements**   The authors would like to thank Oscar Nierstrasz for helpful comments on earlier drafts. The authors would also like thank the various anonymous reviewers of earlier drafts of this paper. They have certainly helped to improve this paper, and the authors are indebted to their care and consideration.

---

[12] http://ecs.victoria.ac.nz/~djp/files/QuickCheckForWhiley-2020-09-27.zip

**Finding Bugs with Specification-Based Testing is Easy!**

## A  Experimental Data

**Finding Bugs with Specification-Based Testing is Easy!**

■ **Table 3** Performance results for benchmarks with differing scopes and sampling rates. Mean execution time (in seconds) is reported across 10 actual runs with 5 warmup runs discarded. A timeout of 1200 s was employed for the 15 runs of each benchmark and sampling / scope configuration. Missing values indicate timeouts occurred.

| Benchmark | tiny/s | small/s | medium/s | | large/s | | | huge/s | | | |
|---|---|---|---|---|---|---|---|---|---|---|---|
|  | 1.0 | 1.0 | 0.1 | 1.0 | 0.01 | 0.1 | 1.0 | 0.001 | 0.01 | 0.1 | 1.0 |
| 001_average | 0.0 | 0.0 | 0.0 | 0.0 | 0.02 | 0.03 | 0.02 | 0.06 | 0.06 | 0.06 | 0.45 |
| 002_fib | 0.0 | 0.0 | 0.0 | 0.0 | 0.0 | 0.0 | 0.0 | 0.0 | 0.0 | 0.0 | 0.0 |
| 003_gcd | 0.0 | 0.0 | 0.0 | 0.0 | 0.01 | 0.01 | 0.01 | 0.01 | 0.01 | 0.01 | 0.01 |
| 004_matrix | 0.0 | 0.14 | 19.06 | - | - | - | - | - | - | - | - |
| 006_queens | 0.0 | 0.04 | 0.65 | 5.05 | 35.39 | - | - | - | - | - | - |
| 007_regex | 0.0 | 0.12 | 2.47 | 8.32 | 11.88 | - | - | - | - | - | - |
| 008_scc | 0.0 | 2.31 | - | - | - | - | - | - | - | - | - |
| 009_lz77 | 0.0 | 0.03 | 0.56 | 0.55 | 6.63 | 6.7 | 11.48 | 16.75 | 17.25 | 35.17 | - |
| 010_sort | 0.0 | 0.07 | 1.73 | 1.73 | 23.8 | 23.87 | 26.26 | 37.08 | 36.51 | 46.49 | - |
| 011_codejam | 0.0 | 0.01 | 0.05 | 0.05 | 0.79 | 0.78 | 0.78 | 1.69 | 1.67 | 1.85 | 17.97 |
| 012_cyclic | 0.0 | 0.06 | 3.66 | 3.63 | 23.14 | 23.23 | - | 34.01 | 54.96 | 60.54 | - |
| 013_btree | 0.0 | 0.01 | 8.42 | 8.31 | - | - | - | - | - | - | - |
| 014_lights | 0.0 | 0.04 | 0.04 | 0.04 | 0.04 | 0.04 | 0.04 | 0.04 | 0.04 | 0.04 | 0.04 |
| 015_cashtill | 0.0 | 0.01 | 0.06 | 0.06 | 0.47 | 0.49 | 0.47 | 5.39 | 5.4 | 5.41 | 5.39 |
| 016_date | 0.0 | 0.07 | 0.47 | 0.47 | 1.48 | 1.5 | 1.46 | 3.38 | 3.39 | 3.39 | 3.39 |
| 017_math | 0.01 | 0.04 | 0.64 | 0.64 | 12.25 | 12.2 | 12.27 | 64.98 | 65.12 | 65.17 | - |
| 018_heap | 0.0 | 0.0 | 0.05 | 0.05 | 0.17 | 0.17 | 1.11 | 0.01 | 0.02 | 0.03 | 10.28 |
| 022_cars | 0.0 | 0.01 | 0.04 | 0.04 | 0.11 | 0.1 | 0.11 | 0.22 | 0.22 | 0.22 | 0.22 |
| 023_microwave | 0.0 | 0.01 | 0.01 | 0.01 | 0.01 | 0.01 | 0.01 | 0.02 | 0.02 | 0.02 | 0.02 |
| 024_bits | 0.0 | 0.0 | 0.01 | 0.01 | 0.03 | 0.03 | 0.03 | 0.09 | 0.09 | 0.09 | 0.09 |
| 025_tries | 0.0 | 0.05 | 11.27 | 16.41 | 71.7 | - | - | - | - | - | - |
| 026_reverse | 0.0 | 0.01 | 0.5 | 0.5 | 14.16 | 14.15 | 14.85 | 61.04 | 61.0 | 64.98 | - |
| 027_c_string | 0.0 | 0.0 | 0.06 | 0.06 | 0.46 | 0.47 | 3.28 | 1.02 | 4.14 | 37.57 | - |
| 028_flag | 0.0 | 0.01 | 0.04 | 0.04 | 0.21 | 0.2 | 0.2 | 0.88 | 0.89 | 0.89 | 0.9 |
| 029_bipmatch | 0.0 | 0.03 | 0.03 | 0.53 | 0.22 | 1.34 | 20.91 | 5.34 | 15.19 | 15.83 | 23.02 |
| 030_fracions | 0.0 | 0.01 | 0.06 | 0.06 | 0.28 | 0.27 | 0.27 | 0.61 | 0.62 | 0.62 | 0.8 |
| 032_arrlist | 0.0 | 0.28 | 8.05 | 11.03 | 8.88 | 32.72 | - | 37.52 | - | - | - |
| 033_bank | 0.0 | 0.04 | 0.06 | 0.06 | 0.08 | 0.08 | 0.08 | 0.11 | 0.11 | 0.11 | 0.11 |
| 102_conway | 0.0 | 0.28 | 4.87 | 4.89 | 3.87 | 9.13 | - | 12.74 | 75.27 | - | - |
| 104_tictactoe | 0.01 | 0.01 | 0.02 | 0.02 | 0.02 | 0.02 | 0.02 | 0.02 | 0.02 | 0.02 | 0.02 |
| 107_minesweeper | 0.0 | 0.5 | 14.06 | 32.88 | 12.37 | 17.82 | - | 13.77 | 27.55 | - | - |

■ **Table 4** Precision results for benchmarks illustrating, for each benchmark and scope/sampling configuration, the percentage of mutants identified as erroneous.

| Benchmark | tiny | small | medium | | large | | | huge | | | |
|---|---|---|---|---|---|---|---|---|---|---|---|
| | 1.0 | 1.0 | 0.1 | 1.0 | 0.01 | 0.1 | 1.0 | 0.001 | 0.01 | 0.1 | 1.0 |
| 001_average | 15 % | 53 % | 53 % | 53 % | 53 % | 53 % | 53 % | 46 % | 46 % | 38 % | 53 % |
| 002_fib | 0 % | 18 % | 63 % | 63 % | 63 % | 63 % | 63 % | 72 % | 72 % | 72 % | 72 % |
| 003_gcd | 0 % | 35 % | 35 % | 35 % | 35 % | 35 % | 35 % | 35 % | 35 % | 35 % | 35 % |
| 004_matrix | 0 % | 58 % | 60 % | 42 % | 7 % | 4 % | 5 % | 11 % | 2 % | 2 % | 2 % |
| 006_queens | 0 % | 32 % | 38 % | 38 % | 38 % | 38 % | 38 % | 38 % | 38 % | 38 % | 38 % |
| 007_regex | 0 % | 32 % | 35 % | 35 % | 35 % | 27 % | 20 % | 27 % | 20 % | 20 % | 20 % |
| 008_scc | 6 % | 56 % | 25 % | 23 % | 5 % | 5 % | 3 % | 2 % | 2 % | 2 % | 2 % |
| 009_lz77 | 26 % | 41 % | 51 % | 51 % | 51 % | 51 % | 51 % | 51 % | 51 % | 50 % | - |
| 010_sort | 5 % | 14 % | 54 % | 54 % | 59 % | 59 % | 57 % | 61 % | 62 % | 63 % | 3 % |
| 011_codejam | 11 % | 100 % | 100 % | 100 % | 100 % | 100 % | 100 % | 100 % | 100 % | 100 % | 100 % |
| 012_cyclic | 2 % | 32 % | 54 % | 54 % | 56 % | 56 % | 12 % | 56 % | 54 % | 18 % | - |
| 013_btree | 0 % | 13 % | 48 % | 48 % | 22 % | 22 % | 22 % | 22 % | 22 % | 22 % | 22 % |
| 014_lights | 13 % | 66 % | 66 % | 66 % | 66 % | 66 % | 66 % | 66 % | 66 % | 66 % | 66 % |
| 015_cashtill | 8 % | 10 % | 10 % | 10 % | 10 % | 10 % | 10 % | 9 % | 10 % | 9 % | 9 % |
| 016_date | 0 % | 2 % | 4 % | 4 % | 4 % | 4 % | 4 % | 5 % | 5 % | 5 % | 5 % |
| 017_math | 36 % | 67 % | 71 % | 71 % | 71 % | 71 % | 71 % | 27 % | 27 % | 27 % | 7 % |
| 018_heap | 10 % | 55 % | 63 % | 63 % | 72 % | 73 % | 73 % | 96 % | 97 % | 97 % | 48 % |
| 022_cars | 0 % | 45 % | 45 % | 45 % | 45 % | 45 % | 45 % | 55 % | 55 % | 55 % | 55 % |
| 023_microwave | 0 % | 50 % | 50 % | 50 % | 50 % | 50 % | 50 % | 50 % | 50 % | 50 % | 50 % |
| 024_bits | 64 % | 84 % | 86 % | 86 % | 86 % | 86 % | 86 % | 86 % | 86 % | 86 % | 86 % |
| 025_tries | 9 % | 62 % | 71 % | 71 % | 43 % | 39 % | 10 % | 9 % | 9 % | 9 % | 9 % |
| 026_reverse | 33 % | 73 % | 76 % | 76 % | 76 % | 76 % | 76 % | 67 % | 66 % | 67 % | - |
| 027_c_string | 10 % | 56 % | 72 % | 72 % | 72 % | 72 % | 72 % | 72 % | 72 % | 72 % | 37 % |
| 028_flag | 4 % | 66 % | 88 % | 88 % | 90 % | 90 % | 90 % | 90 % | 90 % | 90 % | 90 % |
| 029_bipmatch | 0 % | 64 % | 38 % | 76 % | 34 % | - | - | - | - | - | - |
| 030_fractions | 5 % | 80 % | 80 % | 80 % | 80 % | 80 % | 80 % | 80 % | 80 % | 80 % | 80 % |
| 032_arrlist | 0 % | 76 % | 76 % | 76 % | 76 % | 68 % | 5 % | 60 % | 5 % | 5 % | 2 % |
| 033_bank | 0 % | 14 % | 14 % | 14 % | 14 % | 14 % | 14 % | 14 % | 14 % | 14 % | 10 % |
| 102_conway | 1 % | 26 % | 26 % | 26 % | 25 % | 25 % | 2 % | 25 % | 19 % | 9 % | 1 % |
| 104_tictactoe | 13 % | 12 % | 12 % | 12 % | 12 % | 12 % | 12 % | 12 % | 12 % | 12 % | 12 % |
| 107_minesweeper | 0 % | 42 % | 44 % | 44 % | 43 % | 44 % | 2 % | 43 % | 43 % | 2 % | - |

**Finding Bugs with Specification-Based Testing is Easy!**

**About the authors**

**Janice Chin** is a former student at Victoria University of Wellington, now working as Software Engineer at Harmonic Analytics Limited. Contact her at chinjani@myvuw.ac.nz.

**David J. Pearce** is an Associate Professor of Computer Science at Victoria University of Wellington, and author of the Whiley programming language. Contact him at david.pearce@ecs.vuw.ac.nz.